# When Proximity Falls Short: Inequalities in Commuting and Accessibility by Public Transport in Santiago, Chile

Cesar Marin-Flores[1], Leo Ferres[2,3] and Henrikki Tenkanen[1]

[1]GIST Lab, Department of Built Environment, School of Engineering, Aalto University, Finland
[2] Data Science Institute, Universidad del Desarrollo, Chile
[3] Institute for Scientific Interchange, Turin, Italy

## Abstract

Traditional measures of urban accessibility often rely on static models or survey data. However, location information from mobile networks enables large-scale, dynamic analyses of how people navigate cities. In this study, we employ eXtended Detail Records (XDRs) from mobile phone activity to analyze commuting patterns and accessibility inequalities in Santiago, Chile. We identify residential and work locations and model commuting routes by public transport and walking using the R5 multimodal routing engine. Spatial patterns are examined using bivariate local indicators of spatial autocorrelation (LISA) alongside regression techniques to identify distinct commuting behaviors and their alignment with vulnerable population groups. Our results show that while average public transport commuting times do not differ significantly across socioeconomic groups, marked inequalities emerge when accessibility is considered. High-income neighborhoods consistently exhibit high accessibility, whereas low-income areas show substantially lower levels. Importantly, these disparities do not translate into longer commuting times for lower-income groups, indicating a weak relationship between proximity to opportunities and observed travel times. The analysis also reveals significant disparities across sociodemographic groups, particularly in relation to Indigenous populations and gender. The proposed approach is readily scalable and can support evaluations of changes in commuting patterns and the impacts of urban interventions.

## 1. Introduction

Urban mobility has traditionally been studied using data sources like travel surveys. While these conventional sources provide valuable insights, they also have important limitations: surveys are expensive, infrequent, and often fail to capture spatial-temporal variations in travel patterns (Wang, He and Leung, 2018). In recent years, the UN Statistical Commission has highlighted the potential of mobile phone data to complement official statistics, offering more timely, continuous, and cost-effective information on population movement, although additional work is still needed to fully unlock this potential (United Nations Statistical Commission, 2025). Among these emerging sources, eXtended Detail Records (XDR) stand out as a particularly promising option. XDRs contain all billable events generated by users providing a scalable and inexpensive source of mobility information. This type of data has already been applied successfully in areas such as transportation (Chin *et al.*, 2019) and human migration (Aydoğdu and Salah, 2024; Elejalde *et al.*, 2024).

Despite their long-standing role in mobility research, traditional travel surveys face a key methodological limitation. Because they rely on self-reported information, they are vulnerable to recall errors and respondent fatigue, both of which increase the likelihood of misreporting (Shen and Stopher, 2014). Comparative studies using mobile phone and GPS data consistently show that conventional travel diaries under report trips, in some cases by more than 30 percent of all movements (Bricka and Bhat, 2006; Stopher and Greaves, 2009; Stopher and Shen, 2011). In addition, their high cost and logistical complexity constrain sample sizes and limit how frequently they can be conducted, which reduces their ability to monitor short-term and medium-term changes in mobility patterns. While traditional surveys remain essential for capturing socio demographic characteristics and personal perceptions, mobile phone data offers a powerful complement. XDR data enriches these traditional sources with continuous and, population scale observations that address some of their inherent limitations.

In this work, we combine XDR data with information from traditional surveys to contextualize observed mobility patterns across the urban landscape. This enables us to report accessibility conditions and inequalities in a way that is sensitive to temporal change: our methodological approach could provide near real-time accessibility measurements whenever mobile phone data are available. This offers a level of temporal responsiveness that traditional methods alone cannot achieve, particularly in Chile, where such data currently have gaps of more than ten years.

To complement the interpretation of these mobility patterns, we draw on the concept of accessibility. First introduced by Hansen (1959), accessibility refers to the potential opportunities different groups of people can reach, usually measured by distance, cost, or travel time. Integrating mobility data within an accessibility framework offers researchers a novel perspective to better understand urban systems (Järv *et al.*, 2018; Abbiasov *et al.*, 2024). We use this combined approach to move beyond abstract measures of potential access, toward a more grounded understanding of how people actually interact with urban infrastructure in their daily lives (Leite Rodrigues *et al.*, 2021).

Due to its frequency and non-discretionary nature, commuting is a central focus for studying accessibility. The time devoted to commute tends to come at the expense of secondary activities and can therefore affect people's well-being (Allen *et al.*, 2022). Given its central role in shaping urban life, commuting has been the subject of extensive research, particularly in relation to how it reflects and reinforces social inequalities (Frederick and Gilderbloom, 2018; Allen *et al.*, 2022). However, much of this literature focuses on cities in the Global North, where high-quality data and well-established research infrastructures support ongoing analysis (Schwanen, 2017; Tenkanen *et al.*, 2023). In contrast, studies examining cities in the Global South remain relatively scarce, even though these urban contexts often exhibit more pronounced spatial and social inequalities. This gap is especially evident in research utilizing big data sources, which can capture complex—and often informal—mobility experiences of residents in rapidly growing and unequal cities (Björkegren, 2020).

Identifying neighborhoods that are disadvantaged in terms of accessibility or commuting has long been a central concern in urban research, because it provides the foundation for developing targeted transport and land use policies (Nicoletti, Sirenko and Verma, 2023; Jang and Park, 2025). In research fields like geographical information science (GIScience), spatial statistical methods such as Local Indicators of Spatial Association (LISA) are commonly used to detect those geographical patterns (Anselin, 1995). These approaches typically focus on identifying where inequalities occur, providing limited insight into the composition of the clusters or the underlying factors driving these disparities. To address this limitation, we begin with a descriptive analysis to characterize the demographic composition of the spatial clusters. This step then guides a more formal investigation, where we build on recent work that integrates LISA with regression models (Adorno, Pereira and Amaral, 2025). In our case, we combine a bivariate LISA with multinomial logistic regression not to only map spatial clusters but also to investigate the demographic characteristics of the populations within them.

This study explores how socioeconomic inequalities shape commuting patterns in Santiago de Chile by combining accessibility analysis with mobile phone data. The analysis is guided by three main research questions: 1) How do accessibility measures compare with observed commuting patterns derived from XDR data across different areas of the city and socioeconomic groups? 2) Are there specific areas in the city where high or low commuting times coincide with high or low economic status? 3) How are demographic factors associated with disparities in accessibility and commuting, and which population groups are most affected or advantaged by these inequalities?

To answer these questions, we first estimate average commuting times and access to opportunities using mobile phone data and routing algorithms. We then apply a bivariate local spatial autocorrelation analysis (LISA) to identify spatial clusters where commuting time and the social-material index align. Finally, we characterize these clusters using statistical tests and multinomial logistic regression to assess differences across population groups. This comprehensive approach allows us to understand how overlapping social and economic inequalities shape access to opportunities and the lived commuting experience in the city.

## 2. Background and related research

### 2.1 Human mobility studies based on mobile phone data

In recent years, the increasing availability of massive geospatial data has opened new opportunities for analyzing human mobility patterns. Early contributions anticipated the use of mobile phone data to study space-time behaviors in society (Ahas and Mark, 2005), highlighting its potential as a novel tool to understand urban dynamics (Ratti, C. *et al.*, 2006). Building on this foundation, researchers have extended the use of mobile phone data to diverse urban contexts: from analyzing behavioral changes during emergencies and special events (Calabrese *et al.*, 2010; Bagrow, Wang and Barabási, 2011; Lu, Bengtsson and Holme, 2012), to studying individual trajectories and migration patterns (Blumenstock, 2012; Schaub, 2012). A significant part of this literature demonstrates how mobile phone data can be used to estimate commuting flows in urban environments at both individual and aggregated levels (González, Hidalgo and Barabási, 2008; Tatem *et al.*, 2009). This enables applied analyses, such as identifying spatial mismatches between housing and employment (Zhou, Chen and Zhang, 2016) or using it as an indicator of urban economic activity (Kreindler and Miyauchi, 2023).

Mobile phone data (MPD) has since become central to mobility studies, offering unmatched temporal granularity, wide population coverage, and greater cost-efficiency compared to traditional sources like travel surveys or censuses (Palmer *et al.*, 2013; Calabrese, Ferrari and Blondel, 2015). More recently, the use of eXtended Detail Records (XDRs) has gained prominence. Unlike Call Detail Records (CDRs), which are limited to voice calls and SMS activity, XDRs capture data usage events, enabling significantly higher temporal resolution and allowing for more continuous tracking of mobility behaviors (Calabrese, Ferrari and Blondel, 2015). Because of these advantages, XDRs have become especially useful to construct detailed mobility trajectories, facilitating the inference of modes of transportation (Graells-Garrido *et al.*, 2023) and enhancing home location algorithms (Pappalardo *et al.*, 2021).

Although mobile phone data has become increasingly granular and continuous, especially through XDRs, spatial precision is still a concern in mobility analysis. The spatial resolution of mobile phone data is influenced by the density and distribution of cellular antennas, which varies between urban and rural areas (Calabrese, Ferrari and Blondel, 2015). In urban settings, a higher concentration of antennas allows for more accurate user localization, while sparsely covered areas lead to coarser spatial estimates. The location of mobile devices is typically inferred using Voronoi tessellation (Bonnetain *et al.*, 2019; Sotomayor-Gómez and Samaniego, 2020). While the exact position within a Voronoi polygon is often unspecified, this method helps to mitigate spatial heterogeneity by defining more consistent coverage zones. To further reduce biases from uneven antenna distribution and to create more uniform grids, techniques such as dasymetric interpolation have been applied (Järv, Tenkanen and Toivonen, 2017), refining spatial units for more reliable mobility comparisons.

### 2.2 Mobility approach to reveal accessibility disparities

The concept of accessibility has long been central to urban studies and transportation research. Early definitions describe it as the ease with which people can interact with places (Hansen, 1959). This was later expanded to emphasize individuals' freedom to reach opportunities distributed across urban space (Kwan, 1998). In this sense, accessibility reflects the spatial and temporal potential for individuals to engage with places. This idea is reinforced by the notion that areas with high accessibility enhance the potential for participation in desired activities (Wang and Mu, 2018). Recent work has further highlighted the importance of individual constraints—such as income, time budgets, and transport affordability—that influence one's ability to access opportunities (Tiznado Aitken, Palm and Farber, 2024). These constraints are particularly relevant in cities marked by inequality and informality, where they can significantly limit freedom of access.

In cities of the Global South, where urban expansion often outpaces the development of formal infrastructure, accessibility is shaped not only by the availability of transportation but also by persistent socioeconomic inequalities (Moreno-Monroy, Lovelace and Ramos, 2018). In São Paulo, for example, studies have shown that during weekdays, residents of low-income areas travel farther on average to reach key activity centers compared to those from more affluent neighborhoods (Leite Rodrigues *et al.*, 2021). Similarly, in Mexico City, it was found that although lower-income workers often face longer commutes, they tend to limit their mobility to reduce transportation costs, sometimes opting for informal employment opportunities closer to home (Suárez, Murata and Delgado Campos, 2016). In this context, disparities in accessibility across the city highlight how urban mobility patterns are shaped as much by necessity as by choice (Moreno-Monroy and Posada, 2018).

To operationalize and measure accessibility inequalities, researchers have employed various distributional tools. The Lorenz curve and Gini coefficient, for instance, have been used for decades to examine disparities in access to healthcare, employment, and green areas (Neutens *et al.*, 2010; Ni *et al.*, 2024; Martin and Conway, 2025). More recently, metrics such as the Palma ratio have gained attention in accessibility studies. Originally introduced in the context of income inequality (Palma, 2011), the Palma ratio compares the average accessibility of the richest 10% with that of the poorest 40%, thereby emphasizing group-based disparities. This measure has been effectively applied in various urban contexts without requiring heavy computational resources (Oviedo and Guzman, 2020; Li *et al.*, 2023), and even at the national scale (Pönkänen, Tenkanen and Mladenović, 2025).

Despite the growing interest in accessibility and the development of new measurement approaches, many studies still rely solely on static data sources, such as modeled travel times or census-based trip matrices (Swanson and Guikema, 2024) . These methods offer only a partial view of actual behavior, often assuming fixed origins and destinations and emphasizing proximity-based access. In contrast, mobile phone data enable dynamic and large-scale measurement of mobility patterns. Recent research has demonstrated the potential of such data to supplement traditional travel surveys by identifying anchor locations, inferring origin-destination (OD) matrices, and even estimating modes of transportation (Pappalardo *et al.*, 2021; Graells-Garrido *et al.*, 2023). The integration of this dynamic, behaviorally informed data into accessibility research (Järv *et al.*, 2018) opens new possibilities for understanding how infrastructure, policy, and socioeconomic factors shape the geography of opportunity in urban environments.

**3. Data and methods**

This study examines the metropolitan area of Santiago, the most populated city in Chile, home to nearly 7 million people. The analysis covers 35 independently administered municipalities. In recent years, the city has undergone rapid demographic expansion, largely driven by both internal and international migration (J. Rodríguez Vignoli, 2019; INE, 2023) . These dynamics have deepened spatial inequalities, making Santiago a central case for analyzing patterns of mobility and access to opportunities (Ropert *et al.*, 2023). To derive key statistics for commuting patterns and to assess related inequalities, we employ a three-step methodology (Figure 1).

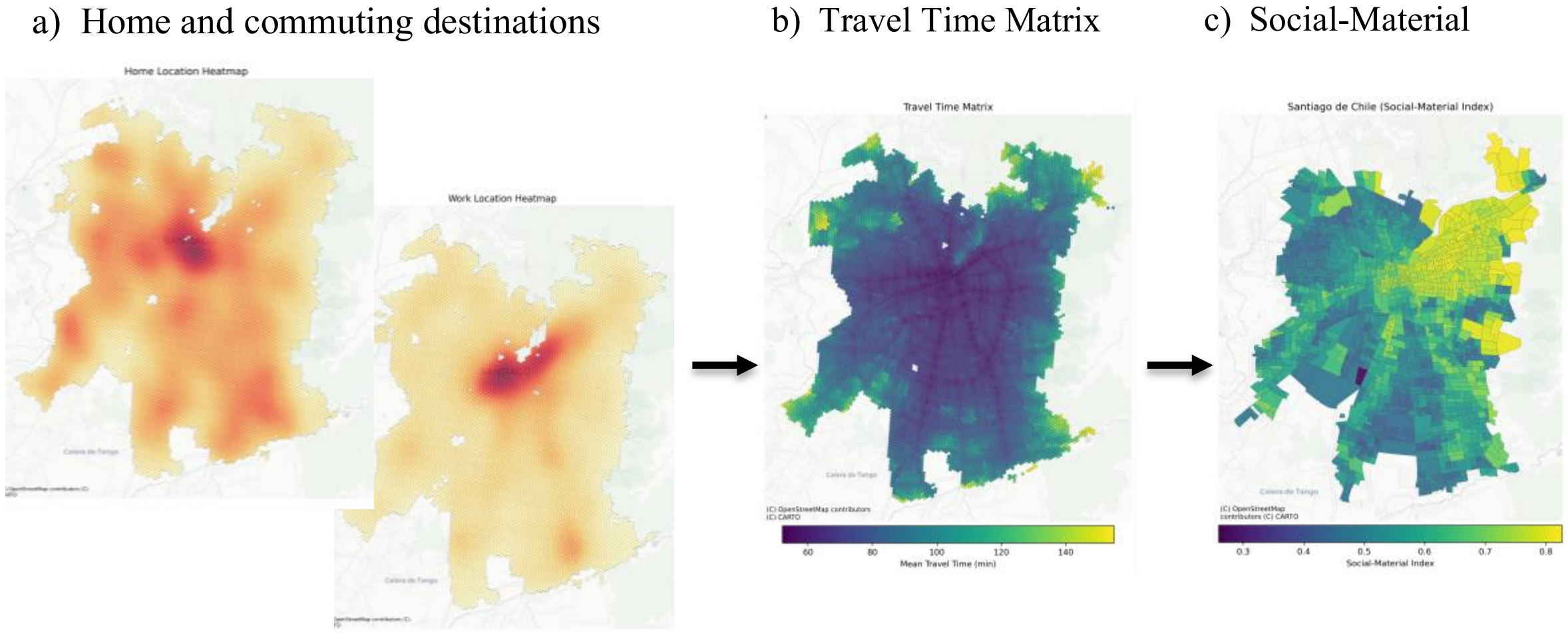

Figure 1. In our study design, we combine information about a) home and key opportunity locations derived from mobile phone data, b) travel times across the city calculated for each H3 Level 9 hexagon, and c) the social-material index obtained from open-source data.

**3.1 eXtended Detail Records data**

We analyzed one month of XDR data from Telefónica Chile, a mobile phone operator covering approximately 28–31% of the national market. The dataset includes all billable events from more than 1,300 Base Transceiver Stations (BTS) distributed across urban Santiago. Each XDR record was aggregated into hourly bins, allowing us to estimate commuting patterns based on the frequency and timing of mobile phone connections. The analyzed data corresponds to March 2023. We restricted our analysis to active monthly users, defined as individuals with an average of more than two signals per day in the studied month. This filter helps to avoid unreliable estimations based on sparse data.

Although the data come from a single operator, previous studies show that Telefónica's mobility data provides a reliable representation of city-wide movement patterns. For instance, anonymized CDRs from the same operator showed a strong correlation ($\rho = 0.89$) with Santiago's official origin–destination survey (Graells-Garrido and Saez-Trumper, 2016), indicating that mobility flows are well captured despite partial market coverage. Similarly, a high correspondence was found between Telefónica's inferred nighttime population and the 2017 Census with $R^2 = 0.96$, suggesting that the operator's user base adequately reflects the spatial distribution of the population and, by extension, captures mobility across different socioeconomic groups, which was the central focus of a study in the context of the COVID-19 lockdown (Gozzi *et al.*, 2021).

### 3.2 Origin-Destination (OD) places of commuting

To identify residential locations, we filtered all signals occurring during nighttime hours (23:00–06:59) and assigned them different weights depending on the hour of the day. The weighting scheme used to infer home and work locations follows time-dependent dynamics reported in previous studies (Zhang *et al.*, 2019; Bergroth *et al.*, 2022), as well as in national time-use data. In Chile, mobile phone data from the same provider as in this study were used to analyze daily journeys and urban dynamics in Santiago (Graells-Garrido and Saez-Trumper, 2016). Their findings show that the number of active weekday trips begins to decline around 19:00, becomes very low after 23:00, and reaches negligible levels between 2:00 and 4:00 a.m. Additionally, the most recent National Time Use Survey (Instituto Nacional de Estadística, 2023) reveals that fewer than 35% of the population remains out of bed at midnight, falling to approximately 5% between 2:00 and 5:00 a.m., before increasing again above 20% after 06:00 a.m. Based on this empirical evidence, we assigned different weights to detect the residential locations: a weight of 2 was used between 00:00 and 02:00 and between 04:00 and 06:00, while a weight of 3 was used between 02:00 and 04:00 reflecting the high likelihood that users are at home during these hours. Figure 2 presents a graphical summary of the survey-based evidence alongside the weighting scheme used for home detection, which is formalized in Equation 1.

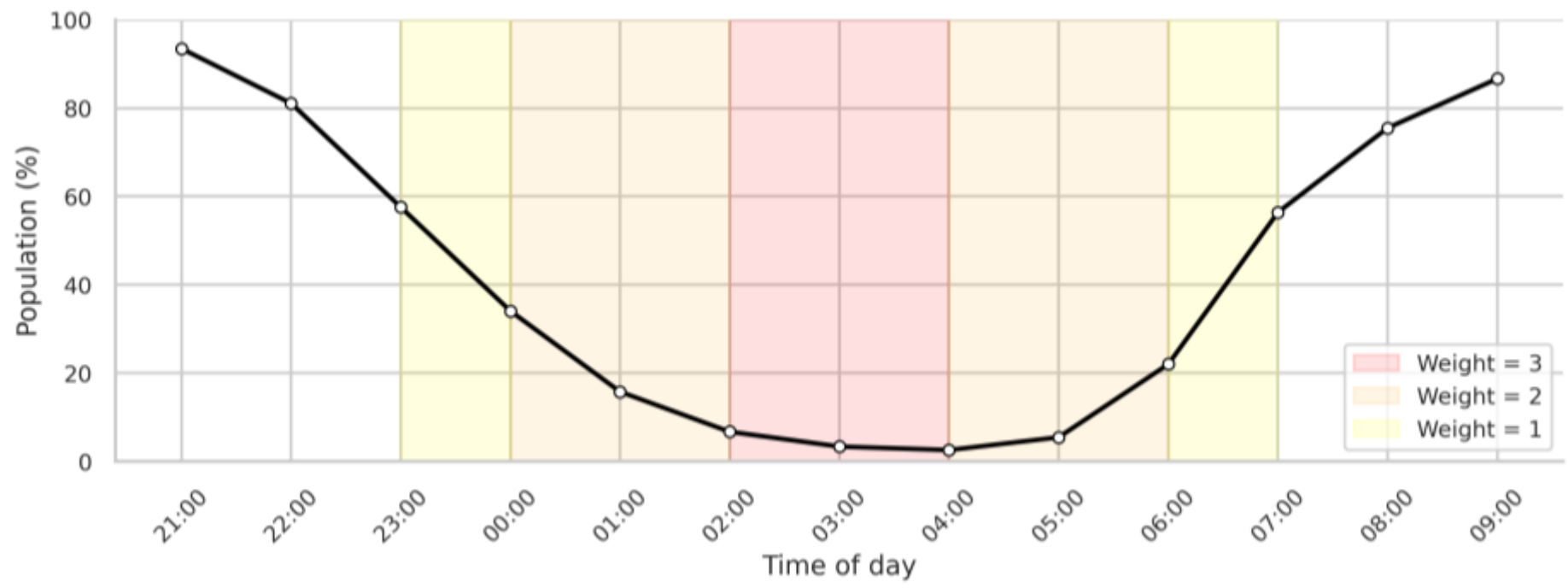

Figure 2. Proportion of the population outside of home during specific hours of the day, based on responses from the National Time Use Survey (Instituto Nacional de Estadistica, 2023). Shaded areas indicate the weighting scheme used to estimate home locations in the analysis: yellow = weight 1, orange = weight 2, red = weight 3.

$$w_h(h) = \begin{cases} 3 \;\; if \;\; h = 2,3 \\ 2 \;\; if \;\; h = 0,1,4,5 \\ 1 \;\; \text{if} \;\; h = 6,23 \\ 0 \;\; otherwise \end{cases} \tag{1}$$

Each phone user *i*, in month *m*, received a total home weight score for each BTS *j*, calculated as:

$$H_{i,j}^{(m)} = \sum_{h \in \text{night}} w_h(h) \cdot N_{i,j,h}^{(m)} \tag{2}$$

where $N_{i,j,h}^{(m)}$ is the number of connections made by user *i* to BTS *j* during hour *h* in month *m*. The BTS with the maximum home score was selected as the home location:

$$\text{home}_i^{(m)} = \arg\max_j H_{i,j}^{(m)} \tag{3}$$

For work locations, a similar time-dependent approach was applied. The weighting scheme was aligned with the main working hours reported in the Time Use Survey, indicating that most formal work activities occur between approximately 09:00 and 18:00. A decrease in weights was applied during typical lunch hours, in accordance with patterns commonly observed in Chilean work schedules (Jara-Díaz, Munizaga and Olguín, 2013). This adjustment ensures that we are capturing the temporal structure of standard work routines in Santiago while minimizing misclassification arising from midday non-work-related movements.

$$w_w(h) = \begin{cases} 2 \;\; if \;\; h = 9, 10, 11, 14, 15, 16, 17 \\ 1 \;\; if \;\; h = 12, 13 \\ 0 \;\; otherwise \end{cases} \tag{4}$$

The corresponding total weight score for each user and BTS was computed as:

$$W_{i,j}^{(m)} = \sum_{h \in \text{work}} w_w(h) \cdot N_{i,j,h}^{(m)} \tag{5}$$

The work location was defined as the BTS with the highest total work score that is not the same as the home location:

$$\text{work}_i^{(m)} = \arg \max_{j \neq \text{home}_i^{(m)}} W_{i,j}^{(m)} \tag{6}$$

This approach produces a monthly origin–destination dataset where each user is associated with a pair of locations ($home^{(m)}$, $work^{(m)}$). We refer to the work-time anchor places (Ahas *et al.*, 2010) as the key opportunities, since they typically represent the most important place where individuals spend time during the day, such as for work or education. The full code used to generate the results is openly available at (REMOVED FROM PEER REVIEW). The data used in this study came from a private provider, one of the major telecom operators in Chile. Due to privacy agreements, the data cannot be shared publicly.

### 3.3 Estimating travel times by public transport and walking

We estimated multimodal travel times using the R5py Python library (Fink *et al.*, 2022), which relies on the R5 routing engine (Conway, Byrd and Van Der Linden, 2017; Conway, Byrd and Van Eggermond, 2018). This tool supports integrated public transport and pedestrian routing, making it well-suited for modeling commuting scenarios. For each origin-destination (OD) pair, we estimated the shortest travel time by simulating departures at every minute during the morning peak period (07:00–09:00). Travel was considered either by walking combined with public transport or by walking alone if it was the fastest travel option. From these simulations, we reported the first percentile travel time, representing the fastest 1% of routes. The routing accounted for transit schedules, service frequencies, and walking connections. The street network data was sourced from OpenStreetMap, and public transport schedules were obtained from open GTFS feeds provided by the city's transit authority.

To spatially define the locations of origins and destinations, we used mobile phone tower (BTS) locations as reference points. Following Deville *et al.* (2014), we generated Voronoi polygons around each BTS to approximate its coverage area (see Figure 3A–B). These polygons were then discretized into a uniform hexagonal grid using Uber's H3 spatial index at resolution level 9. This level was chosen based on the density of BTS towers in urban Santiago, ensuring both high spatial granularity and comparability across different urban zones. As part of the preprocessing, we also removed areas overlapping with national parks or protected lands, as well as polygons where no residents were identified in the home-detection process. This filtering ensured that the final spatial units captured meaningful inhabited urban areas suitable for estimating commuting times.

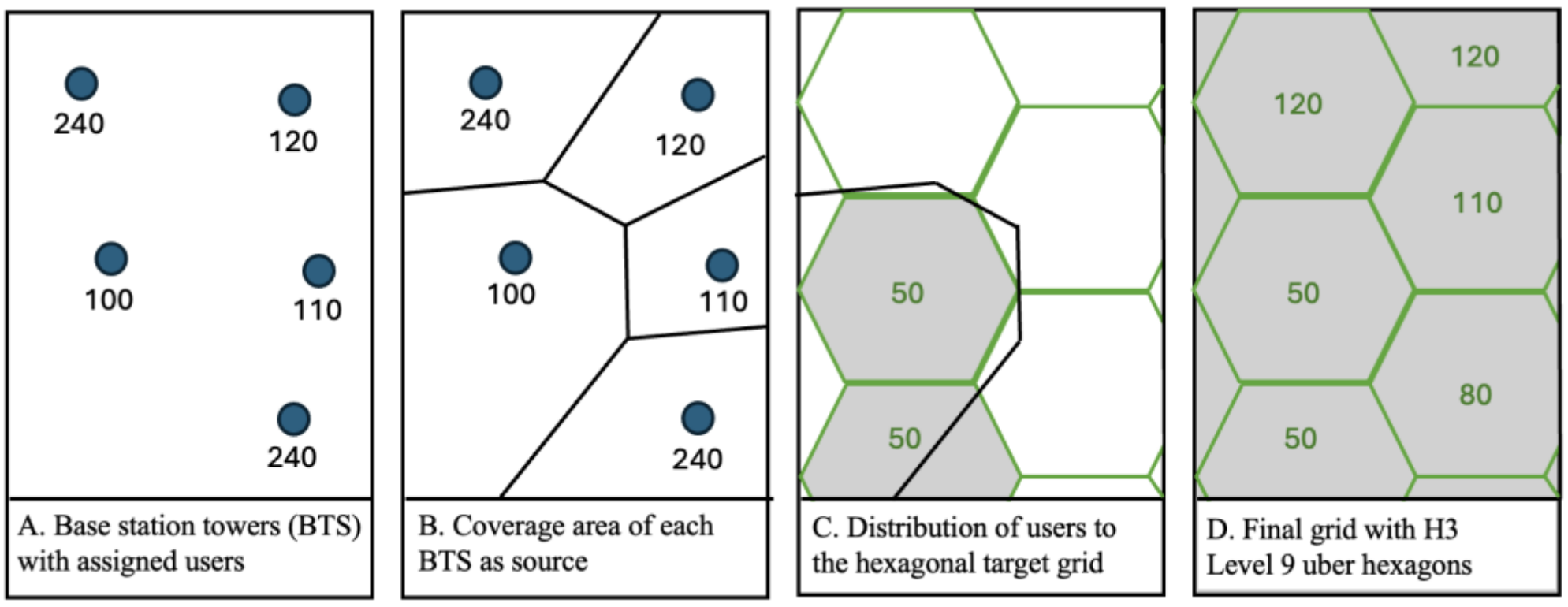

Figure 3. Spatial transformation of BTS-based locations to the Uber H3 level 9 grid.

Each hexagon was assigned to the Voronoi polygon that covered the largest share of its area. To estimate how users were distributed across space in the Uber's H3 grid, we divided the total number of users linked to each BTS equally among the hexagons within its corresponding Voronoi polygon. As illustrated in Figure 3C, users were evenly distributed across the hexagons contained within the studied Voronoi cell, with each hexagon inheriting an equal share of the BTS's users. The centroid of each hexagon served as the representative location for routing and distance estimation. This transformation enabled the calculation of travel times between all observed home and key opportunity hexagon pairs, ultimately resulting in more than million origin–destination (OD) combinations. To assess the robustness of this spatial selection, we conducted a sensitivity analysis (Supplementary Material A) comparing results derived directly from Voronoi polygons from BTS with those obtained from the hexagonal grid. This comparison evaluated differences in spatial user distributions and overall commuting-time measures, showing that the hexagon-based approach preserved the key patterns while providing a more realistic representation in areas where Voronoi polygons are large and unable to reflect commuting times that are representative of the entire covered space.

### 3.4 Combining mobility and socio-economic indicators

In order to incorporate the socioeconomic dimension into the analysis, we used the Social-Material Index (Observatorio de Ciudades, 2022). This index provides fine spatial detail and captures a broad range of conditions beyond income, including education levels, housing quality, and overcrowding. Its comprehensive approach allowed us to better characterize the socioeconomic context of different urban areas.

#### 3.4.1 Palma ratio

To explore disparities in commuting times, we first used the Palma ratio as a global inequality measure. We computed the Palma ratio as the quotient between the average commuting time of the top 10% most privileged population and that of the bottom 40% most disadvantaged population, based on the Social-Material Index. The Palma ratio is defined as:

$$Palma\ Ratio\ = \frac{\overline{T}_{10}}{\overline{T}_{40}} \quad (7)$$

Where $\overline{T}_{10}$ represents the average commuting time for the top 10% and $\overline{T}_{40}$ for the bottom 40%.

### 3.4.2 Accessibility: Cumulative opportunities measure

Building on the previously estimated travel times, we assessed accessibility using a cumulative opportunities measure (Hansen, 1959) with a continuous distance decay function. The key opportunities considered are the work-time anchor places introduced in Section 3.2. Prior research indicates that commuting behavior is best represented by a downward log-logistic (S-shaped) decay, where sensitivity is lower for very short or very long travel times (De Vries, Nijkamp and Rietveld, 2009). Motivated by this, we implemented the logistic decay curve proposed by Bauer and Groneberg (2016), parameterized by the median ($IP$) and standard deviation ($SD$) of observed travel times, as shown in Eq. (9).

Formally, the cumulative opportunities accessibility (COA) was computed as:

$$COA_{oT} = \sum_{d=1}^{n} P_d\ f(t_{od}) \quad (8)$$

where $P_d$ is the number of opportunities at destination *d*, $t_{od}$ is the travel time between origin *o* and destination *d*, and the logistic decay function is defined following Bauer and Groneberg (2016):

$$f(t_{od})\ = \frac{1 + \exp\left(-\frac{IP \cdot \pi}{SD \cdot 3}\right)}{1 + \exp\left(\frac{t_{od} - IP}{SD \cdot 3}\pi\right)} \quad (9)$$

with *IP* denoting the median and SD the standard deviation of the travel time distribution.

## 3.5 Local spatial autocorrelation and cluster composition analysis

In addition to global measures, we examined the local spatial relationship between socioeconomic conditions and commuting times using the bivariate Local Indicator of Spatial Association (LISA) (Anselin, Syabri, and Smirov, 2002). The bivariate LISA identifies areas where high (or low) commuting times are spatially associated with high (or low) socioeconomic conditions in neighboring areas, providing a detailed view of localized inequalities. The LISA analysis was carried out using the PySal-ESDA library (Rey, Sergio and Anselin, Luc, 2007), with the spatial weight matrix constructed based on Queen contiguity.

After identifying spatial clusters of commuting times and socioeconomic conditions, we evaluated how relevant demographic subgroups were distributed across the resulting LISA categories. Based on data availability and the historical marginalization of certain population groups, we selected five demographic variables from the last available national census, conducted in 2018: gender, immigrant status, elderly population (over 65 years), minors (under 18 years), and Indigenous population.

We first visualized the distribution of these variables across LISA clusters and then statistically assessed differences using Kruskal-Wallis and Dunn's post hoc tests. The Kruskal-Wallis test was chosen as a non-parametric alternative to ANOVA, as it does not require the assumption of normality—a necessary adjustment given that some of the studied subgroups do not exhibit normal distributions. Upon detecting significant differences, we applied Dunn's post hoc test to determine which specific LISA categories differed from each other for each subgroup. To control for the family-wise error rate, we adjusted p-values using the Holm-Sidak method.

### 3.6 Socio-demographics and cluster membership: A multinomial model

To further examine the relationship between socio-demographic variables and spatial clustering in Santiago, we applied a multinomial logistic regression model. This method is appropriate when the dependent variable is categorical and unordered, as is the case with the five LISA categories identified in our analysis: High-High (HH), Low-Low (LL), High-Low (HL), Low-High (LH), and Non-Significant (NS). These categories represent distinct spatial clusters based on socio-material conditions and commuting patterns across the city.

Multinomial logistic regression enables us to assess both the direction and the strength of associations between independent variables—such as the percentage of immigrants, retired individuals, minors, and Indigenous populations—and the likelihood of belonging to each LISA category. Unlike simpler methods that merely detect associations, this approach quantifies the relationships through odds ratios. These ratios indicate how much more (or less) likely a given hexagon is to belong to a particular LISA category relative to the Non-Significant (NS) reference group, providing a more nuanced understanding of the spatial distribution of different population groups.

## 4. Results

### 4.1 Average commuting time and Palma ratio in Santiago

The study draws on data from approximately 1.1 million active users in Santiago. From this population, we identified around 610,000 recurrent commuters for the month of study, based on consistent travel behavior patterns. Among these users, the average commuting time, combining public transport and walking, was 46.1 minutes. This result aligns with previous studies (Herrera and Razmilic, 2018), which reported commuting durations in Santiago typically ranging between 40 and 60 minutes, the longest in Chile.

To assess equity in commuting times across the city, we first calculated the Palma ratio using the socio-economic status of each hexagon from the most granular dataset available (Observatorio de Ciudades, 2022). In Santiago, the Palma ratio is 1.01, indicating relatively low inequality: average commuting times in the top 10% most privileged areas were only slightly shorter than those in the bottom 40%. Although values above 1 reflect a degree of inequality—suggesting better access for wealthier groups—the magnitude here points to an even distribution of commuting durations across these two segments. To complement this measure across the full income spectrum, Figure 4 presents the distribution of commuting times by socio-economic decile using boxplots. This visualization shows that the 80–90% decile is the only group that clearly stands out, exhibiting noticeably lower commuting times compared with both higher and lower deciles.

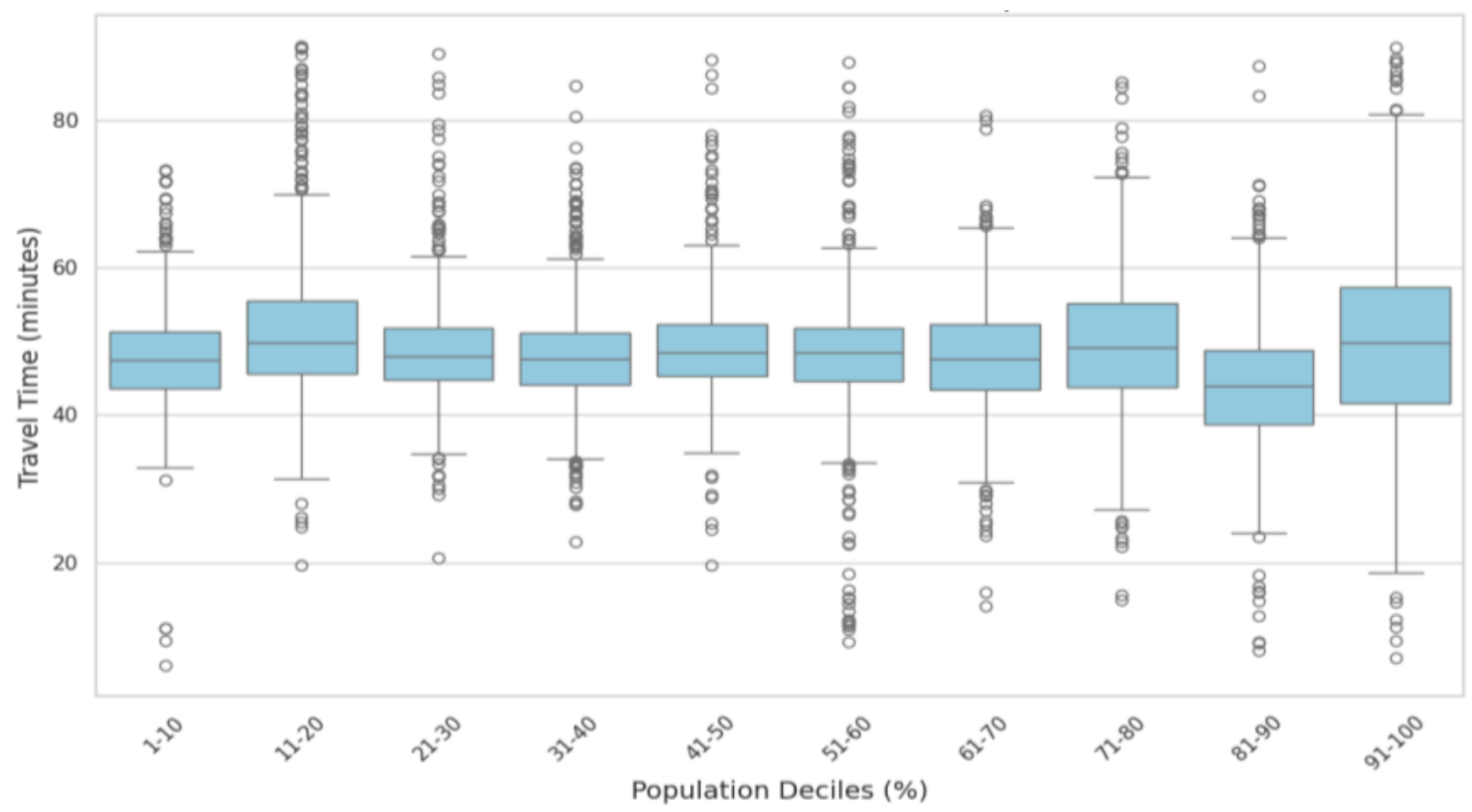

Figure 4. Distribution of travel times across population deciles in Santiago

### 4.2 Spatial patterns of socio-material inequality and commuting time

After identifying overall commuting trends, we examined how these patterns align with the city's socio-economic landscape. A bivariate map (Figure 5) shows the spatial distribution of average commuting time and the Social-Material Index, each divided into quartiles. The northeastern part of Santiago, home to many of the city's most privileged neighborhoods, contains a large concentration of areas in the highest socio-economic quartile (label A in Figure 5). However, this part of the city does not exhibit a big cluster of short commuting times. Only some areas of Providencia, Santiago downtown, Ñuñoa, and Las Condes (label B in Figure 5), where the density of opportunities is particularly high (see Figure 1a), reduced travel durations are observed. The southern periphery displays more defined clusters where social disadvantage and extended commuting times coincide. These patterns are especially evident in municipalities such as La Pintana, Puente Alto, Lo Espejo, and San Bernardo, underscoring persistent spatial inequalities (label C in Figure 5).

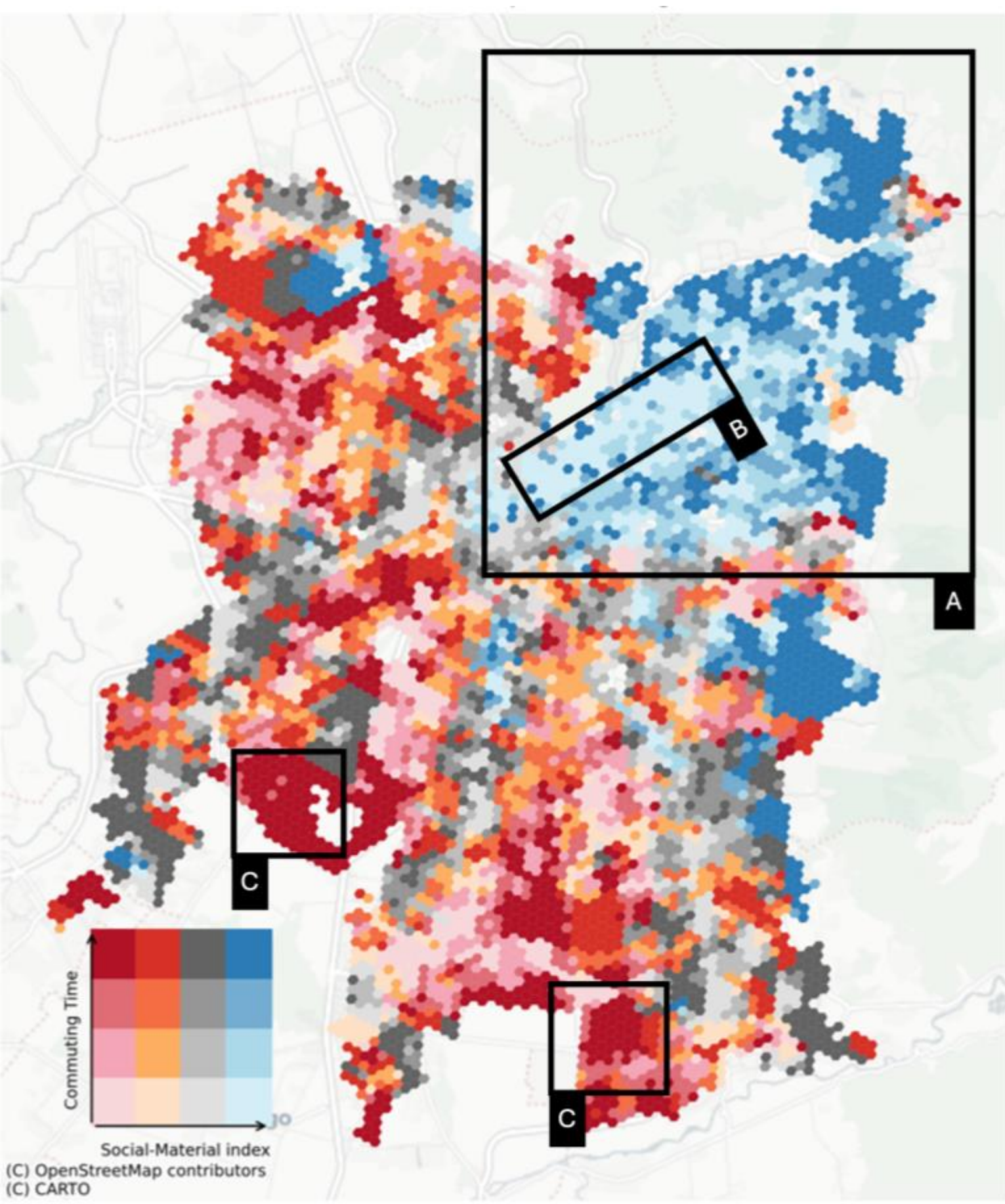

Figure 5. Bivariate map showing the relationship between Social-Material Index and average commuting time (with different colors and hues). Letters A–C mark areas discussed in the text: (A) northeastern sectors with the highest socioeconomic quartile; (B) central areas with dense opportunities and higher-income population; and (C) southern clusters with economic disadvantage and long commuting times.

### 4.3 Inequality in access to opportunities

Figure 6 reveals a long-tailed distribution with a subtle S-shaped pattern. Most data points cluster around the city's average commuting time when accessibility levels are not extreme. Once cumulative accessibility reaches roughly 40% of all opportunities, a moderate decline in average commuting times becomes visible. This pattern indicates that areas with different accessibility levels may still show similar commuting durations, and that only when accessibility becomes exceptionally high do additional opportunities begin to reduce travel times. Overall, this suggests that commuting inequalities are not driven solely by differences in average travel times, but also by the degree of choice people have among the opportunities accessible to them.

A prominent feature of the plot is its long-tail distribution with many areas falling below the 15% mark in terms of accessible opportunities, with a steep drop-off separating them from zones with moderate and high accessibility. This distribution reflects pronounced heterogeneity in accessibility, showing that opportunities are far more accessible for some residents than others. The color gradient, which represents the Social-Material Index (SMI), also illustrates a clear socio-spatial pattern. The hexagons with the highest accessibility, those situated toward the right side of the figure, tend to have higher SMI values, indicating more affluent and resource-rich neighborhoods. In contrast, areas with limited access to opportunities and slightly higher commuting times are often associated with lower SMI scores. This pattern reinforces the idea that accessibility is closely tied to social inequality. These well-served neighborhoods exhibit slightly shorter commuting times, but more importantly, they provide access to a much wider set of reachable destinations. This expanded number of opportunities increases residents' autonomy and enhances their ability to select among valuable services.

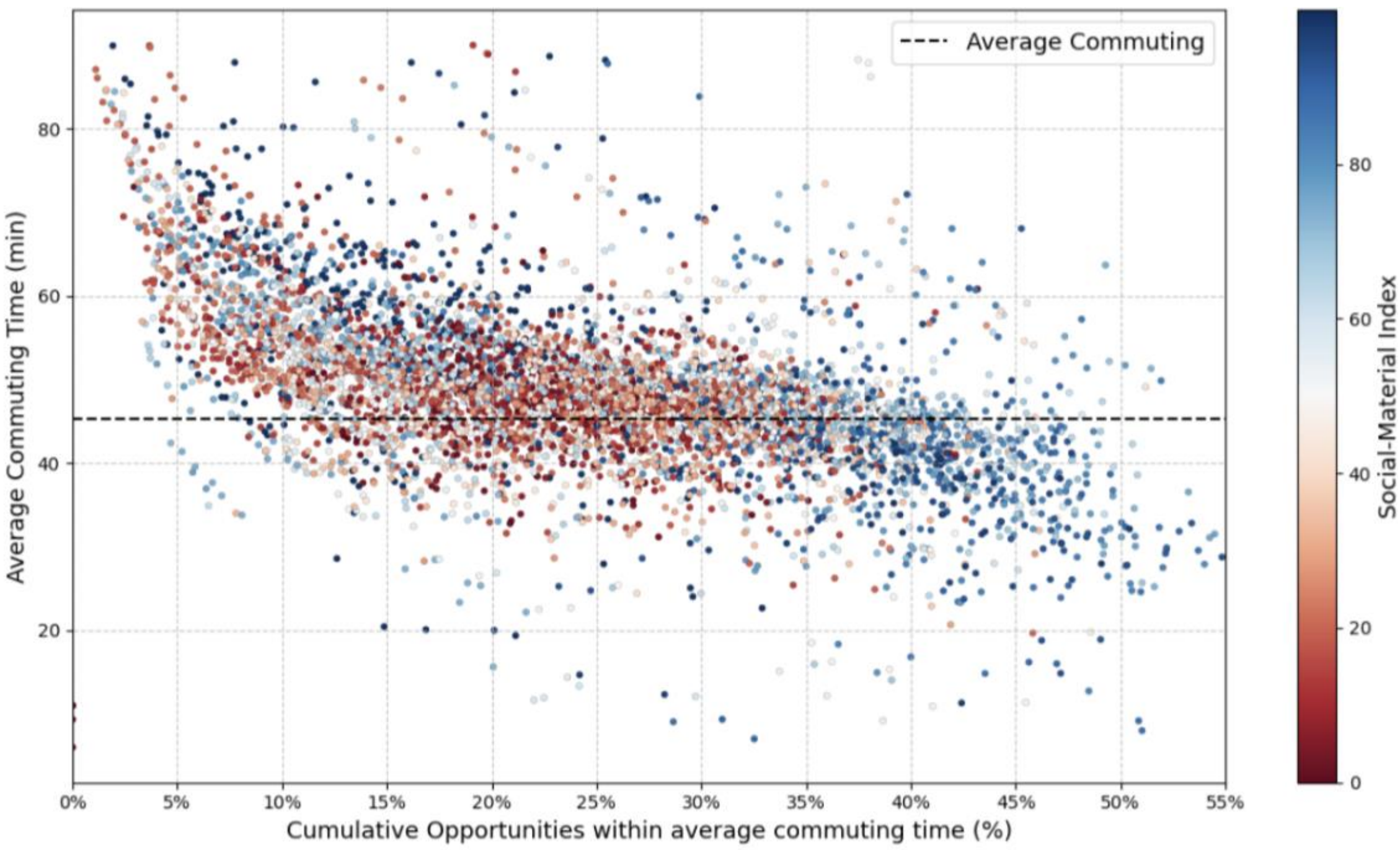

Figure 6. Cumulative key opportunities accessible within the city's average commuting time (46.8 min) from each hexagon (X-axis), compared to the average commuting time of the corresponding area (Y-axis).

## 4.4 Local spatial clusters of inequality

### 4.4.1 Bivariate local indicator of spatial association (LISA)

Figure 7. Bivariate Local indicator of spatial association (LISA) between commuting times and Social-Material Index (SMI).

The Local Indicators of Spatial Association (LISA) analysis (Figure 7) highlights areas in Santiago where commuting time and socio-economic status are significantly locally correlated. In high-income zones, there is only one clear cluster of short commuting times (label C), which corresponds to the Eje Alameda, the city's main commercial and transport corridor. Across the rest of the privileged areas, the map reveals a fragmented pattern, with additional clusters of long commutes appearing only in specific neighborhoods within the municipalities of Lo Barnechea, Peñalolén, and the eastern edge of Las Condes (label A, Figure 7). This spatial heterogeneity suggests that economic affluence does not uniformly translate into reduced travel times by public transport. In lower socio-economic areas, spatial patterns are more pronounced and extensive. Santiago's southern and western peripheries show a consistent alignment between long commuting times and social disadvantage, forming broad zones of compounded vulnerability. Many of these clusters correspond to territories where residential land use coexists with industrial activity as well as informal urban developments known as *poblaciones* (label B, Figure 7)—historically marginalized neighborhoods that still face persistent gaps in access to services and infrastructure. Interestingly, our results reveal that significant clusters were largely absent along Santiago's main transit corridors, including the metro network and other high-capacity public transport routes. This is especially noticeable outside the privileged northeast cluster.

### 4.4.2 Sociodemographic composition of LISA clusters

Building on these findings, we further examine the demographic composition of the LISA bivariate clusters and assess whether specific groups are overrepresented in certain spatial patterns. The five LISA categories include high-high (HH), high-low (HL), low-high (LH), low-low (LL), and not significant (NS). The first letter represents the socio-material index, where "H" indicates high values and "L" indicates low values. The second letter corresponds to commuting times, with "H" denoting long commutes and "L" indicating shorter ones.

Figure 8 presents the distribution of selected sociodemographic groups across the five LISA clusters, revealing important differences in how populations are distributed across economic and mobility-based spatial clusters. The clearest disparity appears in the proportion of indigenous population. Economically advantaged areas, both those with high (HH) and low (HL) commuting times, consistently show lower proportions of indigenous residents compared to the disadvantaged clusters (LL and LH) although the differences between commuting-related groups are not stark. This suggests that while economic segregation aligns with lower indigenous representation in wealthier zones, the presence of indigenous populations is somewhat evenly spread among the economically disadvantaged areas, rather than concentrated in one group alone.

The distributions of retired individuals and minors are more heterogeneous across LISA clusters. Unlike other sociodemographic indicators that often mirror Santiago's well-documented spatial inequalities, these two age-related groups do not follow the expected patterns of segregation by economic status. Moreover, there are no evident differences in their distribution based on commuting time. In contrast, the immigrant population shows more marked separation. The non-significant group displays a wide range of values, with numerous outliers, highlighting the heterogeneous nature of immigrant settlement patterns. The HH and HL clusters, economically advantaged areas, tend to have higher concentrations of immigrants. This is particularly noticeable in the HL group, where short commuting times combine with economic advantage. Finally, gender ratios, measured as the percentage of women in the population, reveal an interesting trend. The HH and HL clusters exhibit a higher median proportion of women, approximately 7% more women than men, compared to other LISA categories, where the gender distribution remains closer to parity.

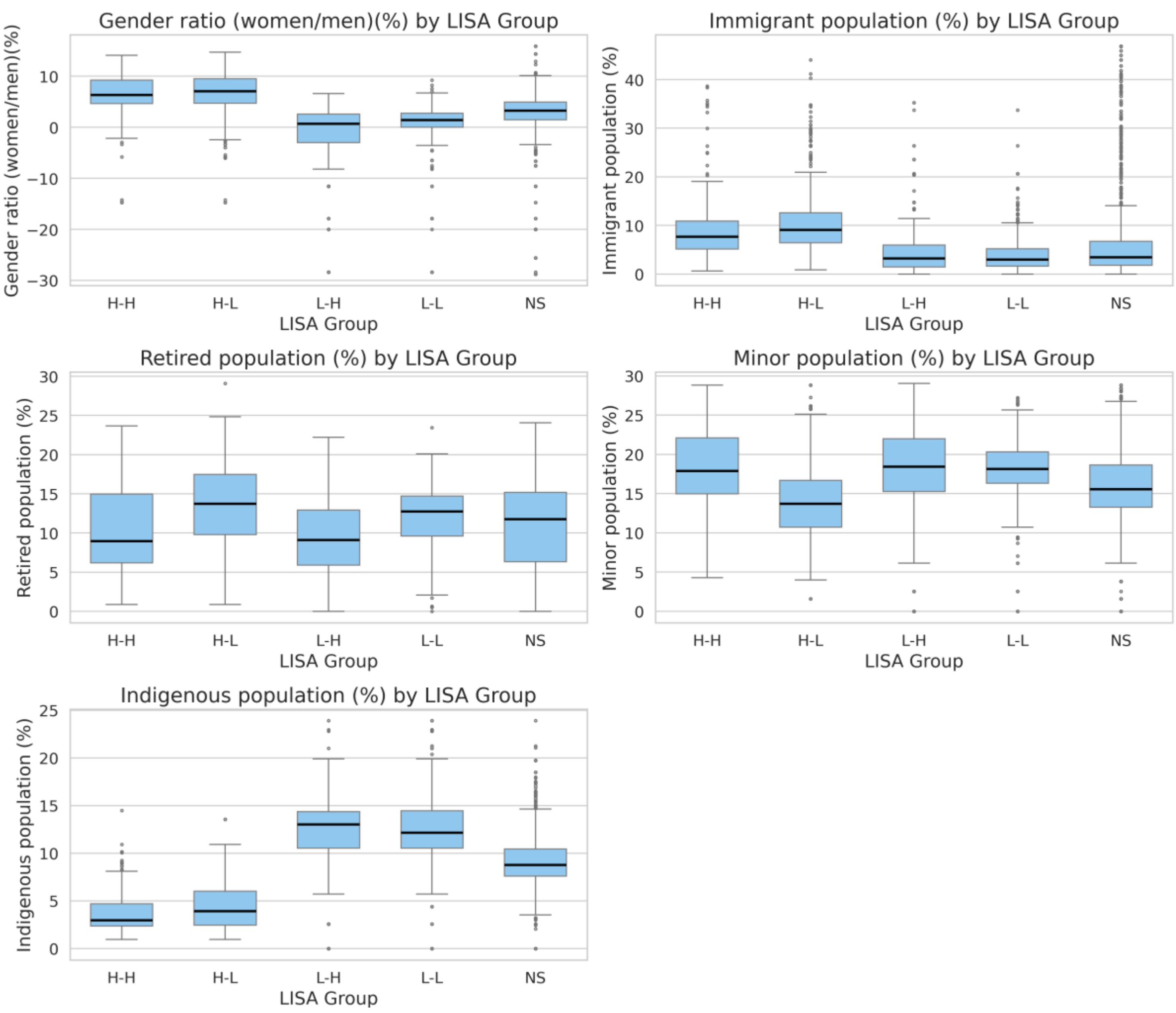

Figure 8. Box plot visualization of sociodemographic groups across LISA cluster types.

To assess whether the observed sociodemographic differences across LISA clusters are statistically significant, we applied the Kruskal–Wallis test. This non-parametric test compares the distributions of multiple independent groups and produces a test statistic called H, which quantifies the degree of difference between groups: a higher H value indicates stronger evidence that at least one group differs from the others. The results, presented in Table 1, show strong evidence of group-level differences, with all p-values well below conventional thresholds ($p < 0.001$). The highest H values were observed for the indigenous population and the gender ratio, suggesting especially pronounced contrasts in their distribution across clusters. These findings reinforce the idea that LISA groupings are not only spatially distinct but also socio-demographically structured.

Table 1. Kruskal–Wallis test results for sociodemographic variables across LISA cluster.

| Sociodemographic groups | Kruskal-Wallis Statistic (H) | p-value |
|---|---|---|
| Gender ratio | 2093.9 | < 0.001 |
| Immigrant population | 1142.9 | < 0.001 |
| Retired population | 234.5 | < 0.001 |
| Minor population | 520.2 | < 0.001 |
| Indigenous population | 3246.7 | < 0.001 |

We gained further insight by applying Dunn's post hoc test (see Supplementary Material B for full results). This analysis showed that most pairwise comparisons between LISA clusters were highly significant, with p-values well below conventional thresholds. The only exceptions were the proportion of the minor population, where the comparison between the HH and LH clusters was not statistically significant ($p \approx 0.2$), and the gender composition, which also showed no statistically significant differences in high-income clusters. This suggests that minors are distributed more evenly across clusters influenced by commuting patterns rather than socio-material factors, and that gender differences are similarly weakly structured in high-income contexts.

### 4.4.3 Multinomial analysis of socio-spatial belonging

Lastly, we used multinomial logistic regression model to explore the association between the socio-demographic groups and spatial cluster membership. The model shows a reasonable performance, with an average classification accuracy of 57% and a McFadden's Pseudo R-squared of 0.42, indicating moderate explanatory power. Table 2 presents the odds ratios for each socio-demographic variable across the five LISA clusters, using the non-significant (NS) group as the reference category. These results highlight distinct patterns of spatial differentiation.

The clearest disparities appear for the indigenous population. In HH clusters, economically advantaged areas with long commutes, the odds of indigenous presence are just 0.024, meaning they are over 25 times less likely to reside there compared to NS areas. Conversely, the odds exceed 5.0 in LL and LH clusters, pointing to a pronounced concentration of indigenous populations in economically disadvantaged areas, regardless of commute length. Gender also shows a notable spatial pattern. In areas with higher proportions of women, the odds ratios for HH and HL clusters exceed 4.0, indicating a strong association between female-dominated populations and economically advantaged zones. This relationship should be interpreted as a statistical correlation rather than a causal effect, as multiple socioeconomic processes, such as household composition, labor force participation, or access to services, may jointly shape these spatial distributions.

The distribution of immigrants is more balanced. In HH and HL clusters, odds ratios are 0.79 and 1.45, indicating a moderate underrepresentation in HH areas and a slight overrepresentation in HL areas, compared to the reference group. However, in LL and LH clusters, the odds drop below 0.45, meaning these areas are less commonly associated with higher proportions of immigrants. This aligns with the broader trend of immigrants settling in more accessible and economically stronger parts of the city.

The patterns observed for retired and minor populations are also notable. Odds ratios for these groups generally fall in a moderate range, indicating that areas with higher shares of retirees or minors are somewhat less likely to belong to economically advantaged clusters. Additionally, both groups show a greater likelihood of being located in LL rather than LH clusters, suggesting that their spatial distribution is correlated not only by economic conditions but also by other factors related to commuting patterns and mobility constraints.

Table 2. Odds ratios from Multinomial Logistic Regression by LISA cluster type.

| LISA Clusters | Gender ratio | Immigrant population | Retired population | Minor population | Indigenous population |
|---|---|---|---|---|---|
| **H-H** | 4.465 | 0.793 | 0.272 | 0.473 | 0.024 |
| **H-L** | 4.455 | 1.448 | 0.344 | 0.207 | 0.056 |
| **L-H** | 0.303 | 0.446 | 3.051 | 2.519 | 5.685 |
| **L-L** | 0.234 | 0.434 | 1.929 | 1.897 | 5.931 |
| **NS** | 1.000 | 1.000 | 1.000 | 1.000 | 1.000 |

## 5. Discussion and conclusions

### 5.1 Inequality in access across different population groups in Santiago

This study investigated commuting patterns in Santiago and their connection to socio-economic factors by combining accessibility measures with observed mobility data. When applying equity measures such as the Palma ratio and examining the full distribution of commuting times across socio-economic deciles, we found that a pattern largely holds: most groups experience comparable commuting durations, with the notable exception of the upper–middle class, which consistently exhibits moderate shorter travel times. These results do not fully align with common assumptions about transport inequality in cities of the Global South, where pronounced spatial and economic divides are often expected to produce large disparities in travel experience (Guzman *et al.*, 2021; De Abreu E Silva and Lucchesi, 2022). In the case of Santiago, our findings suggest that average commuting time alone may not capture the depth of inequality, whereas accessibility-based analyses reveal much stronger socio-spatial disparities across the city.

To better understand the drivers of this apparent parity, we conducted spatial analysis that integrates accessibility to key opportunities with actual travel behavior at high spatial resolution. These opportunities were identified and quantified using mobile phone data, which is particularly valuable in a context where informal job activity represents around 25% of the labor market (INE, 2025), a segment often overlooked in traditional data sources. Mobile phone data tend to offer a more inclusive perspective on commuting and mobility dynamics because of their widespread use across the population, capturing both formal and informal workers. In Chile, mobile penetration is high, with 133 mobile subscriptions per 100 inhabitants (SUBTEL, 2024). Even though we cannot explicitly distinguish between formal and informal workers in our analysis, the underlying data implicitly capture both groups. Moreover, previous research has demonstrated that mobile phone data can effectively capture mobility dynamics in highly informal contexts typically overlooked by conventional surveys (Leite Rodrigues *et al.*, 2021).

Our cumulative opportunities-based accessibility analysis revealed that residents of economically privileged areas consistently have access to a greater number of key opportunities (i.e., work-time anchor locations). However, not all these areas exhibit shorter commuting times; only those with exceptionally high accessibility show noticeably lower travel durations, which further deepens the inequality within this segment. Instead, our results point to a different dimension of commuting inequality: the degree of choice people have within a reasonable travel time. This perspective aligns with the idea of a universal travel time budget, often reported at around 60 to 80 minutes per day (Ahmed and Stopher, 2014) but adds a layer of inequality in terms of what people can access within that time. In privileged areas, this budget allows access to a wide range of opportunities; in less privileged areas, the same budget often limits individuals to a much narrower set of choices. Thus, inequalities may not stem from how long people travel, but by the restricted set of opportunities that certain groups can access within a reasonable travel time.

Our findings also show that the spatial distribution of commuting times does not necessarily follow the well-documented patterns of inequality in Santiago (Larranaga, 2009). However, we identified areas forming large clusters of the most deprived segments, those characterized not only by lower economic status but also by extended travel times. This combination can severely limit residents' ability to access employment and education when travel times or costs exceed reasonable levels. In several of these zones, average commuting durations approach 90 minutes, a threshold often associated with extreme commuting (Allen *et al.*, 2022). Such prolonged travel times have been linked to transport-related time poverty, where excessive commuting reduces the time available for other essential daily activities, including rest, care, and social participation (Tiznado Aitken, Palm and Farber, 2024). This pattern suggests that investments in public transport, especially rail systems, can serve as effective tools to mitigate spatial inequality by improving access to employment and services. Our findings align with evidence from multiple large US cities, where transit investment has been shown to reduce both income inequality and poverty rates, primarily through the expansion of rail networks that complement existing transit systems (Liu and Zhao, 2023).

After identifying distinct spatial patterns of commuting and economic status, a crucial question remained: who are the people behind these patterns? This demographic dimension is often missing in spatial and mobility research, despite its importance in understanding the social structure of cities. From the five population groups examined in this study—indigenous people, immigrants, retirees, minors, and women—clear differences emerged in how these groups are distributed across spatial clusters. One notable insight is the role of immigration in shaping urban settlement patterns. Over the past two decades, migration has significantly contributed to Santiago's population growth and urban transformation. We observed that immigrants concentrate more in areas that offer better accessibility and economic opportunity, suggesting that job prospects and connectivity play a central role in their location choices. Beyond immigration, the spatial distribution of other groups, such as women and indigenous populations, also points to deeper socio-spatial dynamics, with gendered patterns of urban settlement and the persistence of structural exclusion. These findings emphasize that accessibility and commuting inequalities in Santiago are not only about material conditions but are deeply correlated with demographic characteristics and long-standing social hierarchies.

### 5.2 Limitations of the study

Our study includes some limitations that should be acknowledged. First, we do not account for modal choice, as our analysis focuses exclusively on accessibility via the combination of walking and public transport. Travel times by private car are not evaluated, even though car ownership has been reported to be unevenly distributed across the city. Ownership tends to be higher in wealthier areas and is generally associated with shorter commuting times. As a result, the apparent parity in global commuting metrics based solely on public transport may overlook disparities that emerge when other modes are considered. Given that car ownership remains a privilege for many, public transport and walking constitute the primary mobility options for most residents. For this reason, the analysis concentrates on these modes due to their broader social relevance and their critical role in advancing sustainable and equitable cities.

Second, while XDR data provide extensive spatial–temporal coverage, there are limitations. The dataset used in our study comes from a single mobile operator, which may underrepresent certain demographic groups or device types. Although previous studies show that Telefónica's inferred mobility patterns strongly correlate with official travel surveys and census population distributions (Graells-Garrido and Saez-Trumper, 2016; Gozzi *et al.*, 2021), some biases could persist in specific polygons with lower market penetration. Moreover, XDR data capture only billable events, meaning that users with minimal mobile activity may have incomplete trajectories. We mitigated this by excluding low-activity users for the study month, but this necessarily omits individuals whose digital traces are sparse.

Third, our analysis relies on demographic data from the 2017 census, the most recent official dataset available at the time of writing. This limits our ability to assess the impact of more recent demographic changes, particularly migration waves that may have reshaped the mobility patterns of this population in recent years. However, for the rest of the studied population groups, we believe this bias is partially mitigated by the long-term spatial stability of socioeconomic segregation in Santiago, which has remained relatively consistent over the past decades(Fernández, Manuel-Navarrete and Torres-Salinas, 2016) .

Finally, due to privacy concerns and data protection standards, the mobile phone data used in this study presents certain limitations. It cannot be openly shared or accessed, which restricts opportunities for replication or further research using the same dataset. Moreover, all socio-demographic information was anonymized or removed, which prevented us from directly linking individual mobility patterns to user characteristics. As a result, we relied on census-based socio-demographic data assigned according to estimated home locations. This generalization may introduce classification errors, particularly in areas with mixed or changing population profiles.

### 5.3 Future Research

This study opens multiple avenues for further research. Our methodological framework could be expanded to incorporate sustainable transport modes such as cycling, which are increasingly relevant for promoting fair and just accessibility. Integrating active travel into combined accessibility assessments would offer a more holistic understanding of commuting patterns, particularly within the framework of accessibility within planetary boundaries (Willberg *et al.*, 2024). Our analysis shows that only a few areas in Santiago currently achieve commuting times under 15 or even 30 minutes by public transport, highlighting substantial room for improvement and a situation far from the ideal of the 15-minute city concept (Moreno *et al.*, 2021). Future studies could also examine how multimodal combinations, such as cycling to transit hubs, might enhance accessibility in underserved neighborhoods, particularly where direct public transport connections are limited.

Our research relies primarily on mobile phone data to examine the realized use of key opportunities. This provided valuable behavioral information but does not allow us to directly capture the perceptual dimensions of accessibility. Perceived accessibility refers to how individuals subjectively evaluate their ability to reach opportunities based on their own cognitive, social and affective contexts, rather than on purely spatial or infrastructural conditions (Pot, Van Wee and Tillema, 2021). This subjective dimension often diverges from what traditional potential-based accessibility indicators suggest (Lättman, Olsson and Friman, 2018). Consequently, our analysis cannot account for perceptual factors such as feelings of safety, perceptions of transport quality or time pressures, all of which may be particularly relevant in a city like Santiago, where personal experiences can vary substantially across neighborhoods and urban areas. Therefore, we view perception-based approaches such as surveys, participatory mapping and other qualitative methods as important future complements to our work, as they would help connect objective accessibility patterns with the subjective experiences.

In this study, accessibility was continuously assessed based on the number of reachable opportunities defined as the key work-time location where people spend time. However, this approach overlooks important dimensions related to the diversity, quality, and spatial distribution of these destinations. Future research could expand this work by incorporating additional data sources, such as detailed land-use information or Points of Interest (POI) datasets. This would support a more detailed understanding of opportunity types. For example, researchers could distinguish between employment sectors and job characteristics and examine how these relate to the actual availability of meaningful opportunities for residents. Such analysis could help determine whether deprived areas of the city have genuine access to stable and well-paid employment, or whether high opportunity counts hide deeper inequalities in access to specific types of jobs.

### Declaration of generative AI and AI-assisted technologies in the writing process

During the preparation of this work the authors used ChatGPT to improve the English language (no content was generated). After using this tool, the authors reviewed and edited the content as needed and take full responsibility for the content of the publication.

**Supplementary Material A: Sensitivity analysis using Uber H3 Level 9 hexagons vs. Voronoi polygons from Base Tower Stations (BTS)**

To assess the robustness of the results, we compared the commuting time estimates obtained using two different spatial aggregation schemes: Uber H3 Level 9 hexagons and Voronoi polygons derived from Base Tower Station (BTS) coverage areas. At the user level, the average difference in estimated commuting time between the two spatial units was below 2.5%, with mean values of 45.1 minutes for the BTS-based Voronoi polygons and 46.0 minutes for the hexagon-based measures as can be seen in Table 1, indicating that the choice of spatial aggregation has a minimal effect on the overall commuting time results.

Table 1. Commuting times and inequity measures

| Sociodemographic groups | H3 Level 9 | Voronoi polygons |
|---|---|---|
| Average commuting time | 46.0 | 45.1 |
| Palma ratio | 1.01 | 0.95 |

Table 1 also presents the results of the inequity measures discussed in the main text. The concentration index shows only a very small difference, whereas the Palma ratio exhibits the largest gap among these overall metrics, with values of 1.01 for the H3 Uber Level 9 hexagons and 0.95 for the Voronoi polygons. However, this gap remains around 5%.

In Figure 1, we present the bivariate maps described in the main text. Commuting times are indicated by the intensity of the color (lighter tones correspond to shorter times), while the social material index is represented by the color hue. Spatial patterns are consistent across both aggregation units. To complement this analysis, Figure 2 shows the maps after applying the bivariate Local Indicator of Spatial Association (LISA). The patterns observed in the hexagonal units are reproduced in the Voronoi polygons shown on the right side of the figure. Both representations notably reveal a lack of clusters around the metro lines outside the eastern and more affluent parts of the city, as well as the main clusters of long and short commuting times in these high-income areas. The only noticeable difference concerns the most disadvantaged areas at the city's periphery, some of which appear larger in the Voronoi polygons. This likely reflects the larger hexagon sizes in peripheral areas due to the lower density of BTS and reduced population density.

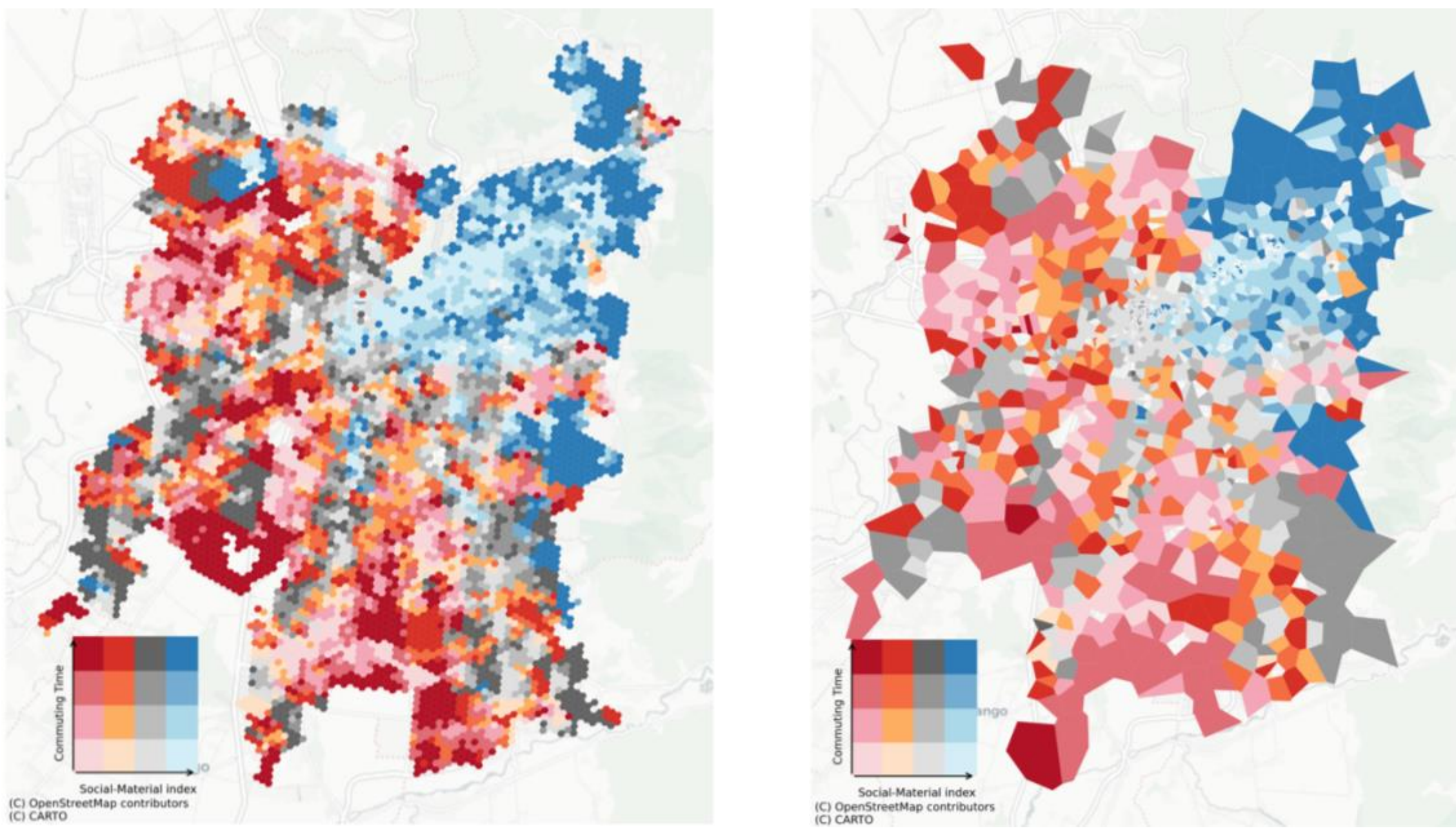

Figure 1. Bivariate maps comparing commuting times and the social material index using Uber H3 Level 9 hexagons (left) and BTS-based Voronoi polygons (right).

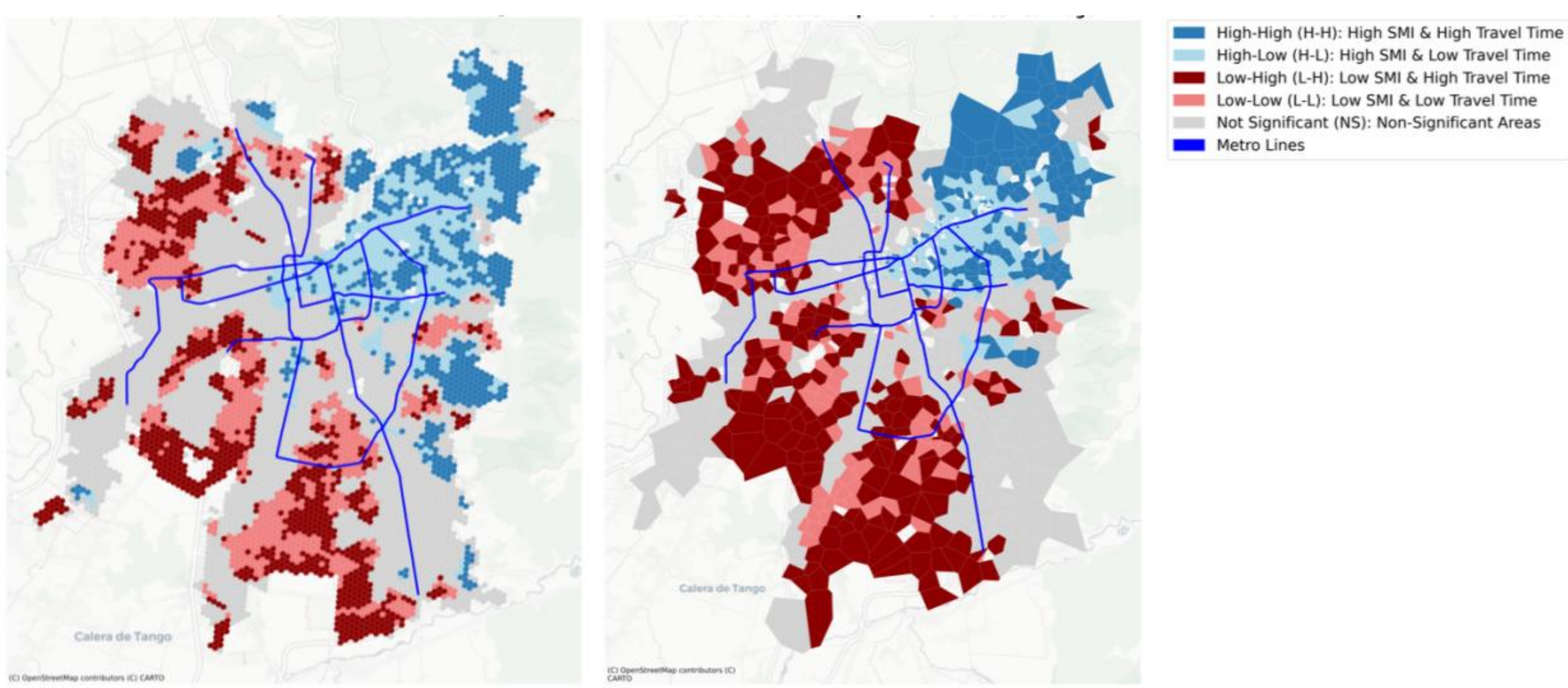

Figure 2. Bivariate Local Indicators of Spatial Association (LISA) between commuting times and the Social-Material Index (SMI). Left: hexagonal units; right: Voronoi polygons.

Figure 3 presents the accessibility and commuting analysis discussed in Section 4.3 of the main text. The number of data points differs considerably between the two aggregation units, with approximately 6,400 hexagons and 1,200 Voronoi polygons, yet the main findings and overall patterns are preserved. The magnitudes along both axes are broadly similar, with cumulative accessible opportunities slightly higher for the Voronoi polygons. The overall pattern of the graph is consistent: economically privileged areas exhibit higher accessibility, and beyond 40% of accessibility this is accompanied by a moderate decrease in commuting times. For areas with medium accessibility, both aggregations show a greater presence of lower-income areas, with commuting times remaining relatively stable in this segment of the graph.

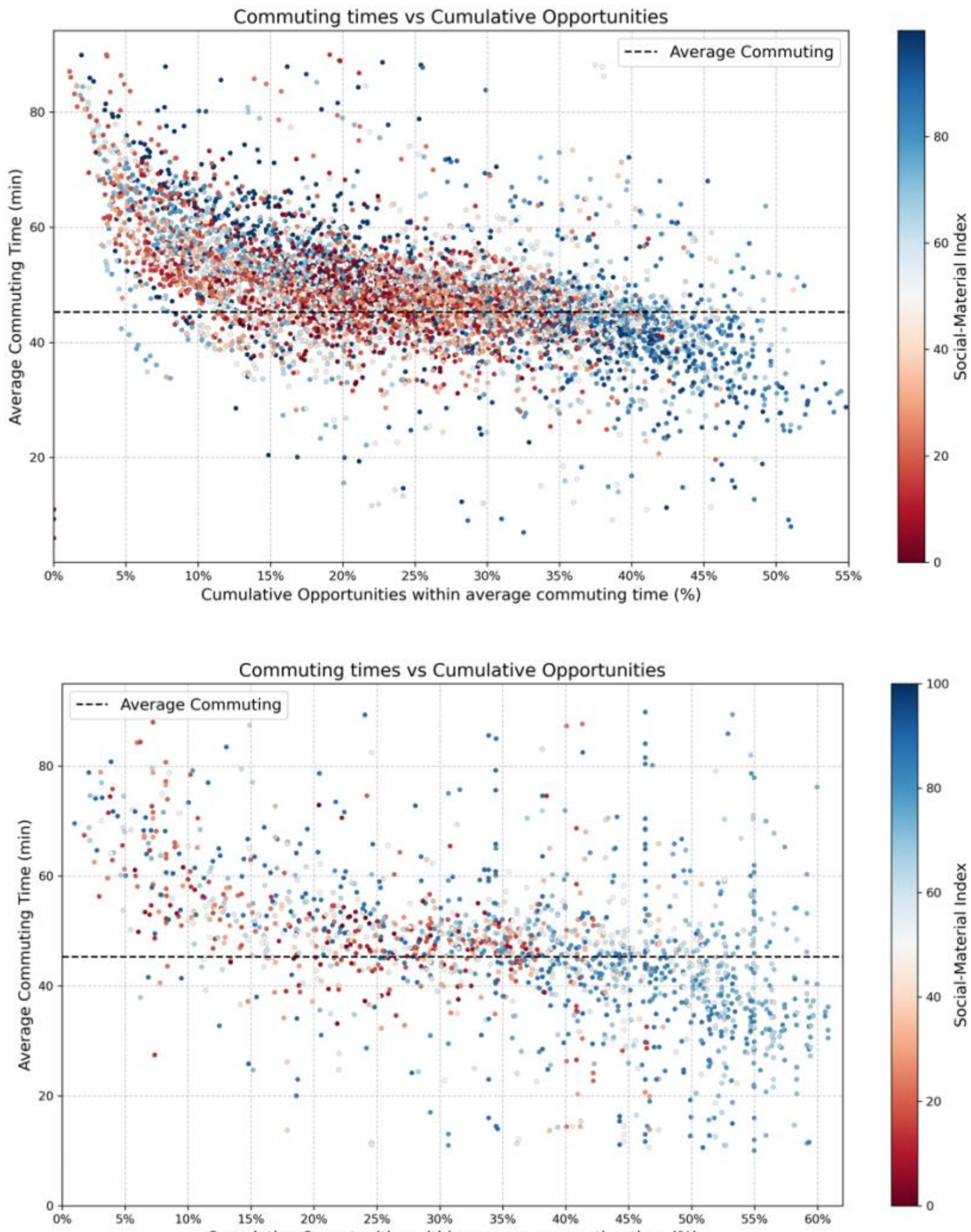

Figure 3. Cumulative key opportunities accessible within the city's average commuting time from each hexagon (X-axis), compared with the average commuting time of the corresponding area (Y-axis). Top: Uber H3 level 9 Hexagons; bottom: Voronoi polygons.

Figure 4 presents an analysis not included in the main text. It shows the percentage difference in commuting times between each hexagon and its spatially corresponding Voronoi polygon. In this comparison, over 63% of the units exhibit an absolute difference of less than 5%, and more than 90% fall below an absolute difference of 10%. The map does not reveal any significant spatial bias in the distribution of errors. As observed in previous figures, the largest differences appear at the outer edges of the city, where hexagons tend to show higher commuting times than their corresponding Voronoi polygons. This likely reflects the larger size of Voronoi polygons in these distant and less densely populated areas, where the centroid of a polygon can diverge substantially from the centroids of the multiple hexagons it encompasses, leading to differences in the estimated commuting times.

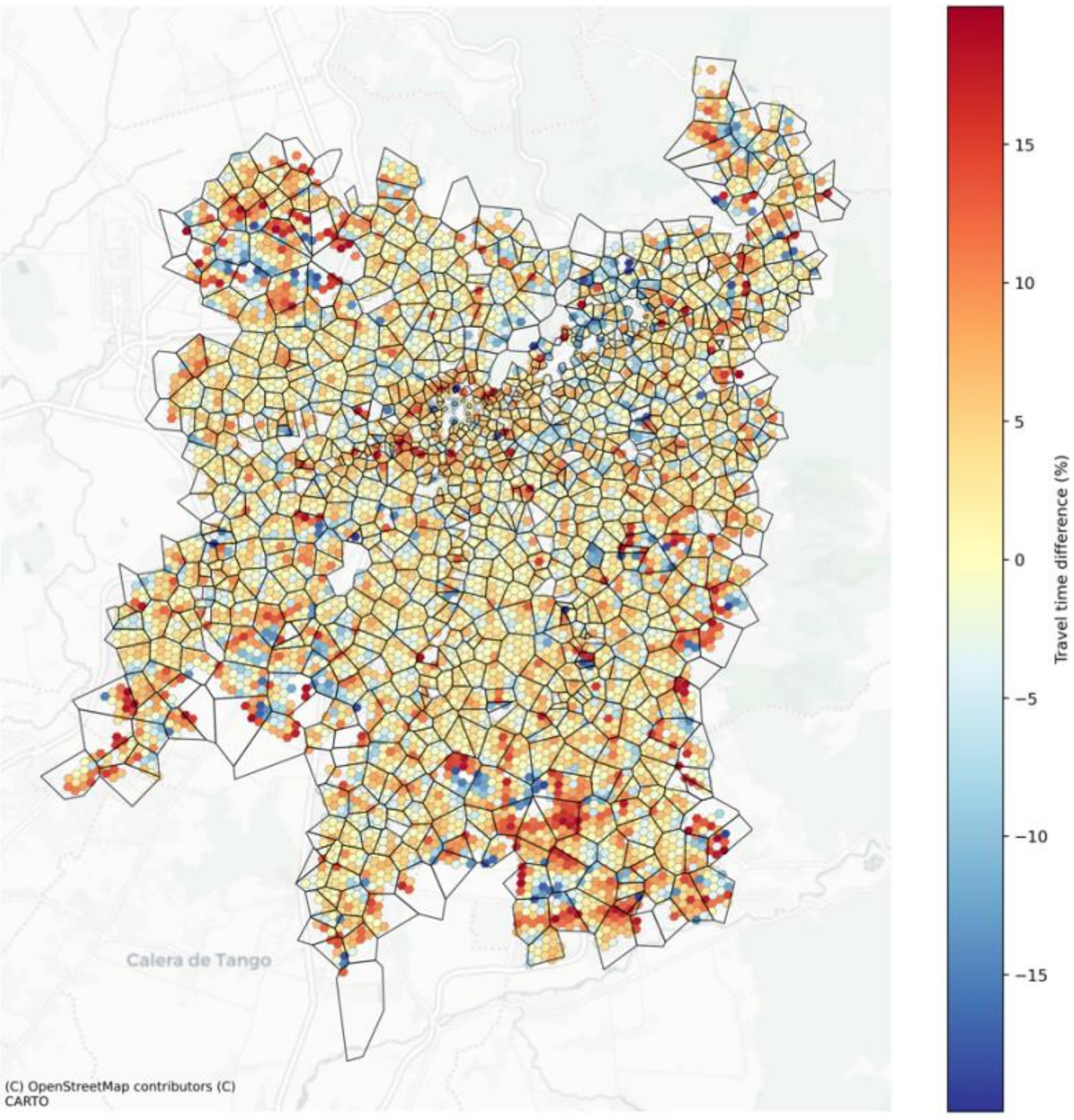

Figure 4**.** Percentage difference in commuting times between each hexagon and its corresponding Voronoi polygon. Voronoi polygon boundaries are overlaid as a reference to illustrate the spatial aggregation.

## Supplementary Material B

Table 1. Dunn’s Test Results for Gender Ratio

| | **H-H** | **H-L** | **L-H** | **L-L** | **NS** |
|---|---|---|---|---|---|
| **H-H** | 1.000 | 0.712 | <0.001 | <0.001 | <0.001 |
| **H-L** | 0.712 | 1.000 | <0.001 | <0.001 | <0.001 |
| **L-H** | <0.001 | <0.001 | 1.000 | <0.001 | <0.001 |
| **L-L** | <0.001 | <0.001 | <0.001 | 1.000 | <0.001 |
| **NS** | <0.001 | <0.001 | <0.001 | <0.001 | 1.000 |

Table 2. Dunn’s Test Results for Immigrant population

| | **H-H** | **H-L** | **L-H** | **L-L** | **NS** |
|---|---|---|---|---|---|
| H-H | 1.000 | 0.055 | <0.001 | <0.001 | <0.001 |
| H-L | 0.055 | 1.000 | <0.001 | <0.001 | <0.001 |
| L-H | <0.001 | <0.001 | 1.000 | 0.092 | <0.001 |
| L-L | <0.001 | <0.001 | 0.092 | 1.000 | <0.001 |
| NS | <0.001 | <0.001 | <0.001 | <0.001 | 1.000 |

Table 3. Dunn’s Test Results for Retired population

| | **H-H** | **H-L** | **L-H** | **L-L** | **NS** |
|---|---|---|---|---|---|
| H-H | 1.000 | <0.001 | <0.001 | <0.001 | <0.001 |
| H-L | <0.001 | 1.000 | <0.001 | <0.001 | <0.001 |
| L-H | <0.001 | <0.001 | 1.000 | <0.001 | <0.001 |
| L-L | <0.001 | <0.001 | <0.001 | 1.000 | <0.001 |
| NS | <0.001 | <0.001 | <0.001 | <0.001 | 1.000 |

Table 4. Dunn’s Test Results for Minor population

| | **H-H** | **H-L** | **L-H** | **L-L** | **NS** |
|---|---|---|---|---|---|
| H-H | 1.000 | <0.001 | 0.228 | 0.212 | <0.001 |
| H-L | <0.001 | 1.000 | <0.001 | <0.001 | <0.001 |
| L-H | 0.228 | <0.001 | 1.000 | 0.011 | <0.001 |
| L-L | 0.212 | <0.001 | 0.011 | 1.000 | <0.001 |
| NS | <0.001 | <0.001 | <0.001 | <0.001 | 1.000 |

Table 5. Dunn’s Test Results for Indigenous population

| | **H-H** | **H-L** | **L-H** | **L-L** | **NS** |
|---|---|---|---|---|---|
| H-H | 1.000 | 0.004 | <0.001 | <0.001 | <0.001 |
| H-L | 0.004 | 1.000 | <0.001 | <0.001 | <0.001 |
| L-H | <0.001 | <0.001 | 1.000 | 0.018 | <0.001 |
| L-L | <0.001 | <0.001 | 0.018 | 1.000 | <0.001 |
| NS | <0.001 | <0.001 | <0.001 | <0.001 | 1.000 |